\documentstyle[prb,aps,epsf]{revtex}
\begin{document}
\twocolumn[\hsize\textwidth\columnwidth\hsize
           \csname @twocolumnfalse\endcsname
%BW5766
\title{Quasiparticle transport equation with collision delay. II.
Microscopic Theory}

\author{V\'aclav \v Spi\v cka and Pavel Lipavsk\'y}
\address{Institute of Physics, Academy of Sciences, Cukrovarnick\'a 10,
16200 Praha 6, Czech Republic}
\author{Klaus Morawetz}
\address{Max-Planck Gesellschaft, Arbeitgruppe ``Theoretische
Vielteilchenphysik'' an der Universit\"at Rostock, D-18055 Rostock,
Germany}
\date{\today}
\maketitle
\begin{abstract}
For a system of non-interacting electrons scattered by neutral
impurities, we derive a modified Boltzmann equation that includes
quasiparticle and virial corrections. We start from quasiclassical
transport equation for non-equilibrium Green's functions and apply
limit of small scattering rates. Resulting transport equation for
quasiparticles has gradient corrections to scattering integrals.
These gradient corrections are rearranged into a form characteristic
for virial corrections.
\end{abstract}
\vskip2pc]
\section{Introduction}
In the first paper of this series\cite{SLM97} (referred as Paper~I),
we have discussed interplay of quasiparticle and virial corrections for
scattering by neutral impurities with resonant levels. We have found
that both corrections are of the same order and tend to mutually
compensate. Accordingly, one should either include both kinds of
corrections or neglect both of them. Separate quasiparticle or separate
virial corrections lead to overestimates of impurity effect on basic
physical quantities like dc conductivity or screening length.

In Paper~I we have used an intuitive modification of the Boltzmann
equation (BE), Eq.~(I-39) [Eq.~(39) of Paper~I]. Long time experience
with quasiparticle corrections shows that intuitive approaches whatever
convincing are far from to be reliable as the wave-function renormalization
factor often emerges in an unexpected way. To become trustworthy,
Eq.~(I-39) has to be recovered from quantum statistics in a very
systematic manner. This is the aim of this paper.

The intuitive modification of the BE we would like to arrive to reads
\begin{eqnarray}
{\partial f\over\partial t}
&+&
z{k\over m}
{\partial f\over\partial r}-
{\partial\phi\over\partial r}
{\partial f\over\partial k}
=-{f\over\tau}
\nonumber\\
&+&
{1\over\tau}{2\pi^2\over k^2}
\int{dp\over(2\pi)^3}
\delta(|p|-|k|)
f(p,r,t-\Delta_t).
\label{qp12}
\end{eqnarray}
This is equation (I-39). It applies to system of non-interacting
electrons scattered by point impurities with resonant levels.
Here, the quasiparticle distribution $f$ is a function of momentum $k$,
coordinate $r$ and time $t$. Lifetime $\tau$, collision delay $\Delta_t$
and wave-function renormalization $z$ are functions of momentum,
potential $\phi$ depends on coordinate and time. The collision delay
$\Delta_t$ (given by the energy derivative of the phase shift)
makes the scattering-in integral non-local in time. This
non-locality represents virial corrections. The quasiparticle
corrections are covered by the wave-function renormalization $z$.

So far, quasiparticle and virial corrections have been studied only
separately using different theoretical tools. Now we briefly review
previous studies to identify which tool is better suited for unified
theory.

\subsection{Virial corrections}
A need to derive virial corrections to a quantum transport equation has
been felt for a while. The progress achieved in this direction can be
represented by Snider's equation.\cite{S90,S91} Using the BBGKY
hierarchy, Snider has derived a quantum transport equation for the
reduced density matrix (Wigner's distribution). Snider's equation is
sufficiently general to describe various corrections beyond the BE,
however, as noticed by Lalo\"e and Mullin,\cite{LM90} all corrections
enter the scattering integral, while quasiparticle corrections should
appear rather as corrections to drift of single-particle excitations.

Snider's equations includes another item alien to the quasiparticle
picture: the reduced density matrix. When the quasiparticle corrections
are in game, the transport equation should deal with the quasiparticle
distribution not with the reduced density matrix. In the reduced density
matrix, contributions from free excitations (quasiparticles) and from
correlations (off-shell motion during dressing processes) are mixed
together. The drift of free excitations can be described by simple
quasiclassical trajectories while correlations require quantum
mechanical treatment. To be able to make efficient approximations,
these two types of motion has to be separated. So far, the theory of
transport based on the reduced density matrix is missing a tool that
would furnish us with such a separation.

\subsection{Quasiparticle corrections}
The second group of papers is a large variety of studies recovering the
Boltzmann equation with quasiparticle corrections in the form visualized
by Landau, e.g.
Refs.~\onlinecite{KB62,E61,NL62,LN62,PK64,PS67,D84,BM90,RS86,SL95}.
On the other hand, none of these studies touches virial corrections. One
expects that the virial corrections should emerge from systematic
quantum-statistical approaches to the Landau theory provided one does
not loose them making unjustified approximations.

>From the intuitive modification of the BE (\ref{qp12}), we can identify
three groups of approximations we have to avoid:
\begin{itemize}
\item
Most of authors limit their attention to the scattering described
within the Born approximation (as we have also done in Ref.
\onlinecite{SL95}). The corresponding T-matrix is just an impurity
potential that does not cause any collision delay. Similarly,
the self-consistent Puff-Whitefield approximation of the electron-phonon
scattering (the Migdal approximation with non-self-consistent phonons)
studied by Prange and Kadanoff,\cite{PK64} gives no virial corrections
since the interaction vertex has no internal electron dynamics like the
T-matrix within the Born approximation.
\item
Prange and Sachs\cite{PS67} have studied the electron-electron and
electron-phonon scattering within the fully self-consistent single-loop
approximation. The screened Coulomb interaction is in general a complex
function of energy so that it yields some kind of virial corrections.
Prange and Sachs have used, however, the static approximation within
which the screened Coulomb potential looses energy dependence and
becomes real. In this way, the virial corrections are lost.
\item
Danielewicz\cite{D84} and Botermans and Malfliet\cite{BM90} have used the
two-particle T-matrix for nucleon-nucleon interaction that definitely
includes virial corrections, however, they have neglected gradient
contributions to the scattering integral. From the non-local character
of the scattering integral in the intuitive BE~(\ref{qp12}),
one can see that the virial corrections are proportional to gradients
of the quasiparticle distribution, see (\ref{qp12a}). Neglecting
gradient contributions one looses the virial corrections.
\end{itemize}
In all the three cases, one can go beyond these approximations. To
recover Eq.~(\ref{qp12}), we have to use the T-matrix for the impurity
scattering and keep the gradient corrections to scattering integrals.
Similar treatment has lead to virial corrections for two-particle
scattering.\cite{MR95}

\subsection{Gradient corrections to scattering integrals}
The gradient corrections to the scattering integral are of central
importance in our treatment. In the intuitive BE (\ref{qp12}), the only
gradient contributions to the scattering integral come from the virial
corrections,
\begin{eqnarray}
{\partial f\over\partial t}
&+&
z{k\over m}
{\partial f\over\partial r}-
{\partial\phi\over\partial r}
{\partial f\over\partial k}
=-{f\over\tau}
\nonumber\\
&+&
{1\over\tau}{2\pi^2\over k^2}
\int{dp\over(2\pi)^3}
\delta(|p|-|k|)
f(p,r,t)
\nonumber\\
&-&
{1\over\tau}{2\pi^2\over k^2}
\int{dp\over(2\pi)^3}
\delta(|p|-|k|)\Delta_t
{\partial\over\partial t} f(p,r,t).
\label{qp12a}
\end{eqnarray}
Accordingly, we expect that gradient corrections resulting within the
standard approaches to the BE are just the virial corrections. Our plan
is to derive gradient corrections from the quasiclassical limit of the
quantum-statistical transport equation and rearrange them into the last
term of Eq.~(\ref{qp12a}).

Gradient corrections to scattering integrals have been already studied
by many authors, the most extensive studies were devoted to the effect
of electric field on collisions. Starting from Barker\cite{B73}, the
field effect on scattering has been discussed for number of scattering
mechanisms within various approaches which include the superoperator
projection technique, Refs.~\onlinecite{B73,BF79}, the Levinson
equation\cite{L69} derived from the BBGKY hierarchy,
Ref.~\onlinecite{CP81}, and non-equilibrium Green's functions,
Refs.~\onlinecite{B81,JW82,JW84,S85,LSV86,KDW87}.

It is important to note that these different approaches resulted in
contradictory predictions. Let us focus on the (linear) gradient
correction to the scattering integral which we claim to be equal to
the virial correction [the last term in the r.h.s of (\ref{qp12a})].
We will assume only self-energies in the Born or Puff-Whitefield
approximation so that the virial corrections are absent, i.e., we
expect the gradient corrections to be absent, too. Surprisingly, this
is not true about gradient corrections one finds in print. One can split
all papers into three groups:
\begin{enumerate}
\item
In Refs.~\onlinecite{B73,BF79,CP81,LSV86},
linear gradient corrections were found.
\item
In Ref.~\onlinecite{KDW87}
linear gradient corrections were found but with opposite sign than
in the first group.
\item
In Refs.~\onlinecite{B81,JW82,JW84,S85}
{\bf no} linear gradient corrections were found.
\end{enumerate}
Unless we will understand why the gradient corrections appear for the
Born or Puff-Whitefield approximation and why there are contradictory
predictions from alternative approaches, we can hardly use the gradient
corrections as a start line of our approach to the virial corrections.

To identify the origin of the three contradictory predictions, let us
inspect what kind of the transport equation is specific for each group.
Although various approaches have been used by authors, we will describe
all the three groups within a common dialect of Green's functions.

\subsubsection{Integro-differential transport equation for the reduced
density matrix}
The integro-differential transport equation for the reduced density
matrix is obtained by the Generalized Kadanoff and Baym (GKB) ansatz
\cite{LSV86} implemented in the time diagonal of the
integro-differential Kadanoff and Baym (KB) equation\cite{KB62}.
>From the GKB ansatz, time argument $t$ of the reduced density matrix
results retarded compared to the time argument $T$ of the scattering
integral, i.e. $t<T$. In the integro-differential equation, it is
natural to identify the instant of the collision with $T$. Using the
linear expansion $\rho(t)=\rho(T)+{\partial\rho\over\partial t}(t-T)$
one then obtains a Boltzmann-like scattering integral from $\rho(T)$ and
gradient corrections from ${\partial\rho\over\partial t}(t-T)$.

\subsubsection{Integral transport equation for the reduced density
matrix}
The integral transport equation for the reduced density matrix is
obtained by the GKB ansatz implemented in the time diagonal of the
integral GKB equation\cite{LW72}. In the integral equation one has a
freedom to identify the instant of the collision with $t$ and gradient
corrections emerge from matching of the scattering integral with the
subsequent propagation. The opposite signs found in groups 1 and 2 thus
follow from the fact that in group 1 authors extrapolate along the
initial state of the collision while in group 2 authors extrapolate
along the final one.

\subsubsection{Integro-differential transport equation for the
quasiparticle distribution}
The transport equation for the quasiparticle distribution is obtained by
the original KB ansatz\cite{KB62} implemented to the KB equation at the
quasiparticle pole. From the KB ansatz, the time argument of the
quasiparticle distribution results identical to the time argument of the
scattering integral. This time is naturally identified as an instant of
the collision event and no gradient corrections appear.

The contradictory predictions of the gradient corrections were found to
follow from different treatments of quasiparticle corrections.
\cite{LKW91}
As shown in
Ref.~\onlinecite{LKW91}, the linear gradient correction to the rate of
scattering out is nothing but the time dependence of the wave-function
renormalization. In other words, gradient and quasiparticle corrections
are linked together. The approach that takes care of quasiparticle
features and naturally leads to the Landau-Boltzmann type of transport
equation for quasiparticles is the one of the group 3, i.e., based on
the KB ansatz. The absence of the gradient corrections for the Born
approximation is also in agreement with our expectation.

\subsection{Non-virial gradient corrections}
Among studies of the gradient corrections that are not devoted to the
field effect we want to point out the paper by Kolomiets and Plyuiko
\cite{KP94}. In the quasiclassical limit, they have evaluated the
scattering integral from the self-energy in the second-order
approximation of the electron-electron interaction keeping all gradient
terms. Similarly to the spirit of classical virial corrections, they
have expressed gradient corrections in terms of effective shifts in
space and momentum. The scattering integral they have derived thus
reminds the one we are looking for.

In spite of formal similarity, the corrections derived by Kolomiets
and Plyuiko are not the virial corrections. Kolomiets and Plyuiko have
used a mixed approach implementing the KB ansatz in the time diagonal
of the integro-differential form of the GKB equation. These two steps
are not compatible as the KB ansatz includes the quasiparticle
distribution while the time diagonal (or integral over energy in
Wigner's representation) provides drift terms for the reduced density
matrix. The gradient contributions they have found are thus a strange
form of the quasiparticle corrections.

Again we came to the same moral. If we want to put equality between the
virial and the gradient corrections to the scattering integral, we have
to work fully in the quasiparticle picture.

\subsection{Presented approach}
Our approach is based on two limits applied to non-equilibrium Green's
functions. The first one is the quasiclassical limit that is inevitable
for the BE.\cite{KB62} The second one is the limit of small scattering
rates.\cite{SL95} This limit restricts validity of quasiparticle
corrections to weakly renormalized systems. The validity of virial
corrections is restricted to the second virial coefficient.

The limit of small scattering rates from the quasiclassical transport
equation for non-equilibrium Green's function has been already presented
in detail in Ref.~\onlinecite{SL95}. In fact, the transport equation for
quasiparticles, Eq.~(70) of Ref.~\onlinecite{SL95}, reduces to the
modified BE (\ref{qp12}). From this point of view, this paper is just an
application of the method derived in Ref.~\onlinecite{SL95}. On the
other hand, in Ref.~\onlinecite{SL95} authors discussed only simple
self-energies on the level of Born approximation which provide
quasiparticle corrections but no virial corrections appear. Accordingly,
there is no collision delay. Here we use a more complex self-energy,
averaged T-matrix approximation, that results in non-trivial virial
corrections similarly as it was done in Ref. \onlinecite{MR95} for
two-particle scattering.

In spite or because of many similar and closely related studies of
gradient contributions to transport equations, we feel that we have to
start our treatment directly from non-equilibrium Green's functions
instead of recalling already achieved results. This is because these
treatments differ in seemingly tiny details that become important as
soon as one wants to keep trace of gradients and quasiparticle
corrections in the same time.

Now we can specify our aim. The precursor of equation (\ref{qp12}) is
the quasiparticle Boltzmann equation in semiconductors, Eq. (70) of
Ref. \onlinecite{SL95}, which reads
\begin{equation}
{\partial f\over\partial t}+
{\partial\varepsilon\over\partial k}
{\partial f\over\partial r}-
{\partial\varepsilon\over\partial r}
{\partial f\over\partial k}
=-z(\gamma_\varepsilon f-\sigma^<_\varepsilon).
\label{i1}
\end{equation}
Except for transport vertex $\sigma^<$, all components are determined
by retarded self-energy $\sigma^R$. First, $\gamma=-2{\rm Im}\sigma^R$
provides inverse lifetime, ${1\over\tau}=z\gamma_\varepsilon$. Second,
the quasiparticle energy is given by $\varepsilon={k^2\over m}+\phi+
{\rm Re}\sigma^R_\varepsilon$. Note that unlike in Paper~I,
$\varepsilon$ includes the potential $\phi$. We made this change of
convention to comply with the convention of Ref.~\onlinecite{SL95}.

It will be easy to show that for homogeneously distributed point
impurities the velocity simplifies to the form in (\ref{qp12}),
${\partial\varepsilon\over\partial k}=z{k\over m}$, and similarly
do the force, ${\partial\varepsilon\over\partial r}=
{\partial\phi\over\partial r}$. Apparently, quasiparticle corrections
are explicitly present in equation (\ref{i1}). The viral corrections,
if present, are hidden in the transport vertex $\sigma^<$. The focus of
our interest thus will be to show that the transport vertex can be
rearranged to the form of the scattering-in integral from (\ref{qp12a}),
\begin{eqnarray}
z\sigma^<_\varepsilon
&=&
{1\over\tau}{2\pi^2\over k^2}
\int{dp\over(2\pi)^3}
\delta(|p|-|k|)
f(p,r,t)
\nonumber\\
&-&
{1\over\tau}{2\pi^2\over k^2}
\int{dp\over(2\pi)^3}
\delta(|p|-|k|)\Delta_t
{\partial\over\partial t}
f(p,r,t).
\label{i2}
\end{eqnarray}

\subsection{Content}
Paper is organized as follows. In Sec.~II, we briefly review the method
of Ref.~\onlinecite{SL95}. We introduce the non-equilibrium Green's
functions, the quasiclassical limit, and the limit of small scattering
rates. With help of these tools the Boltzmann-like transport equation
is derived. In Sec.~III, the self-energy and the transport vertex are
specified within the averaged T-matrix approximation. The most essential
part of our paper is in Sec.~IV, where we evaluate the scattering
integral including its gradient corrections. In Sec.~V, these gradient
corrections are rearranged into a form that is identical to the one
intuitively expected and equation (\ref{qp12}) is recovered. In Sec.~VI,
single-electron observables are valuated. Electron density and current
are discussed in detail. We also discuss the density of energy. Sec.~VII
includes summary. In Appendix we discuss relation between virial and
quasiparticle corrections.

\section{Transport equation}
Our focus of interest is the transport vertex $\sigma^<$. The treatment
of the transport vertex, however, is intimately connected to the
transport equation itself. Therefore we find it profitable to briefly
review derivation of equation (\ref{i1}). This also gives us a room to
introduce a necessary set of equations for non-equilibrium Green's
functions and their relations to components of equation (\ref{i1}).

\subsection{Generalized Kadanoff and Baym equation}
Our starting point is the generalized Kadanoff and Baym
equation\cite{LW72}
\begin{equation}
G^<=G^R\Sigma^<G^A,
\label{gf1}
\end{equation}
that is accompanied by the Dyson equation for retarded (advanced)
propagator\cite{LW72}
\begin{equation}
\left(G_0^{-1}-\Sigma^{R,A}\right)G^{R,A}=1,
\label{gf2}
\end{equation}
where the inverse free-particle propagator reads
\begin{equation}
G_0^{-1}(1,2)=
\left({\partial\over\partial t_1}+
\epsilon\left({1\over i}{\partial\over\partial x_1}\right)-
\phi(x_1,t_1)\right)\delta(1-2).
\label{gf3}
\end{equation}
Here, numbers are cumulative variables $1\equiv (t_1,x_1)$, time
and space. Matrix products include integrations over time and space,
$C=AB$ means $C(1,2)=\int dt_3dx_3A(1,3)B(3,2)$. The $\epsilon$ is the
free-electron kinetic energy and $\phi$ is a potential. Our sign
convention for the correlation function $G^<$ is
\begin{equation}
G^<(1,2)={\rm Tr}\left(\hat\rho\psi^\dagger(2)\psi(1)\right),
\label{gf4}
\end{equation}
where $\hat\rho$ is the grand-canonical averaging operator, and
$\psi^\dagger$ and $\psi$ are creation and anihilation operators,
respectively.

Equations (\ref{gf1}-\ref{gf4}) are general identities and
definitions. The particular physical content of these equations is
specified by self-energy $\Sigma^{R,A}$ and transport vertex $\Sigma^<$.
The transport equation (\ref{i1}) follows from set (\ref{gf1}-\ref{gf3})
with no regards to a particular form of self-energy, except that the
scattering rate connected to $\Sigma^{R,A}$ and $\Sigma^<$ is
supposed to be small in a sense specified below. Moreover,
perturbations in the system of electrons are supposed to be smooth
in time and space so that corresponding gradients are also small.
Thus we leave specification of the self-energy for next section, and
focus on the quasiclassical limit (small gradients) and limit of
small scattering rates of equations (\ref{gf1}-\ref{gf3}).

\subsection{Wigner's representation}
In the quasiclassical limit, all operators are conveniently described in
Wigner's mixed representation
\begin{eqnarray}
a(\omega,k,r,t)
&=&
\int d\tau dy\, {\rm e}^{i\omega\tau-iky}
\nonumber\\
&\times &
A(t+\tau/2,r+y/2,t-\tau/2,r-y/2).
\label{gf5}
\end{eqnarray}
We use convention that lowercase denotes Wigner's transform of
operators denoted by uppercase. This convention is in agreement with
the convention used in Paper~I. Indeed, site-diagonal operators used in
Paper~I (like $\Sigma^{R,A,<}$, $V$ and $T^{R,A}$) in Wigner's
representation are momentum independent and equal to their local
elements $\sigma^{R,A,<}$, $v$ and $t^{R,A}$ introduced in Paper~I.

The transformation (\ref{gf5}) mixes together left and right arguments
of the function $A$, therefore it complicates matrix products.
Keeping only the first gradients in time and coordinates, the matrix
product $C=AB$ in the mixed representation reads
\begin{equation}
c=ab+{i\over 2}[a,b],
\label{gf6}
\end{equation}
where the rectangular brackets denotes Poisson's brackets
\begin{equation}
[a,b](\omega,k,r,t)
=
{\partial a\over\partial\omega}{\partial b\over\partial t}
-{\partial a\over\partial t}{\partial b\over\partial\omega}
-{\partial a\over\partial k}{\partial b\over\partial r}
+{\partial a\over\partial r}{\partial b\over\partial  k}.
\label{gf7}
\end{equation}

\subsection{Propagation of single quasiparticle}
In the quasiclassical limit, one restricts the assumed fields $\phi$ to
those
that vary slowly in time and space. In this case, the field is expected
to have a pronounced effect only on long trajectories of electrons. Such
long trajectories necessarily include number of collisions with
impurities. A propagation between subsequent collisions and individual
collisions themselves happen on a very small time and space scale, thus
they should be nearly the same as in a homogeneous and stationary
potential, i.e., in the absence of the field. There might be also a
small effect of the field on these microscopic scales, however, this
field effect can be handled as a correction linear in gradients of the
field.

Propagators $G^{R,A}$ describe a motion between collisions and also
internal dynamics of collisions. Thus we have to find them up to
linear order of gradients. From the Dyson equation (\ref{gf2}) and
its alternative form $G^{R,A}(G_0^{-1}-\Sigma^{R,A})=1$ one finds that
propagators are free of gradients\cite{KB62}
\begin{equation}
g^{R,A}
=
{1\over\omega-\epsilon-\phi-\sigma^{R,A}}.
\label{gf8}
\end{equation}

Being a complex function,
\begin{equation}
\sigma^{R,A}=\sigma\mp{i\over 2}\gamma,
\label{gf9}
\end{equation}
the self-energy describes two kinds of phenomena. Its imaginary part
${1\over 2}\gamma$ descries scattering out of the state of momentum $k$.
Its real part $\sigma$ renormalizes energy of the single-particle-like
state. The renormalized (quasiparticle) energy is given by a position
of the pole of $G^{R,A}$ at the real axis
\begin{equation}
\varepsilon=\epsilon+\phi+\sigma(\varepsilon).
\label{gf10}
\end{equation}
This quasiparticle energy is an ingredient of the quasiparticle BE
(\ref{i1}).

In the limit of small scattering rates (small $\gamma$ expansion,
$-2{\rm Im}\sigma^R\equiv\gamma\to 0$), the pole of the propagator
sits close to the real axis. Then one approximates the spectral
function
\begin{equation}
a=-2{\rm Im}g^R
\label{gf11}
\end{equation}
by its limiting value
\begin{equation}
a=2\pi z\delta(\omega-\varepsilon)+
\gamma{\rm Re}{1\over(\omega-\epsilon-\phi-i0)^2}.
\label{gf12}
\end{equation}
Within the quasiparticle picture, in the spectral function
$a(\omega,k,r,t)$, the $\delta$ function represents a singular
contribution from quasiparticle state of momentum $k$ (at point $r$ and
time $t$). The second term describes various projections of other
quasiparticle states with energy $\varepsilon(p,r,t)=\omega$ into
momentum $k$, in other words, the second term is an off-pole
contribution. Within the quasiparticle picture, these two parts of
the spectral function has to be treated separately.

The norm $z$ of the $\delta$ function is the wave-function
renormalization
\begin{equation}
z={1\over1-\left.{\partial\sigma(\omega)\over\partial\omega}
\right|_{\omega=\varepsilon}}.
\label{gf13}
\end{equation}

\subsection{Completed collisions}
The generalized Kadanoff and Baym equation (\ref{gf1}) includes all
phenomena we are interested in, but in a hidden form. The simple link
between Green's function formulae and intuitive approach to the BE with
virial corrections is obscured by the fact that these two approaches
deal with different objects. While the correlation function $G^<$
includes both, the pole and off-pole, contributions, the quasiparticle
distribution is related only to the pole part. To recover the BE
equation, one has to separate these two parts.

Within Green's function the pole and off-pole propagation are described
by a single transport equation (\ref{gf1}). In contrast, a diffusion of
particles described by the LHS of the BE applies only to the pole part,
while the off-pole part seems to be missing. In fact, it is not missing.
The off-pole propagation is hidden in scattering integrals and relation
between observables and the quasiparticle distribution.

The possibility to move the off-pole propagation into the scattering
integral follows from a hierarchy of time and space scales in the system
that are inevitable for the theory of Boltzmann type. Diffusion of
quasiparticles is well defined on a hydrodynamical scale that includes
a large number of impurities and a large number of collisions per
particle. On the hydrodynamical scale, there is an appreciable effect
of the field $\phi$ which may cause a transfer of quasiparticles over
long distances. Such a massive changes of the system state is
effectively described with help of a differential transport equation
that balances drift with dissipation. In contrast, individual collisions
happen on a microscopic (local) scale on which the effect of the field
$\phi$ is small so that it can either be neglected or included by
corrections linear in its gradients. Thus, a subdynamics on the
microscopic scale can be integrated through and approximated by
effective scattering rates.

Clearly, to recover the BE we have to separate hydrodynamical and
microscopic scales. This separation is equivalent to the separation
of the pole and off-pole parts based on a non-equilibrium modification
of the expansion in small scattering rates.\cite{SL94,SL95}
Alternatively, the separation of pole and off-pole parts of $G^<$ can
be based on the idea of completed collisions. In this paper we focus on
the microscopic mechanism of collisions, thus we follow the completed
collision approach.

For the purpose of motivation, we assume for a while a homogeneous
system that is diagonal in momentum representation. Let us take a look
on a time-diagonal element of $G^<$, say $G^<(t,t;k)$. In the transport
equation (\ref{gf13}), the transport vertex $\Sigma^<$ represents the
last collision due to which a wave function of electron gained a
component of the momentum $k$. This component can belong either to an
asymptotic state that will form new effective quasiparticle state
or to the off-pole contribution due to some other state $p$. The
asymptotic state is on the energy shell, thus it will live on the time
scale of the quasiparticle life time. The off-pole contribution will
vanish on the scale of a quasiparticle formation time. The latter
is much shorter than the former. To distinguish whether the contribution
to $G^<(t,t;k)$ is of pole or off-pole nature, one can monitor a
vicinity of the time $t$, including a close future, to figure out which
part will survive and which part soon disappears. This procedure
corresponds the approach of Fermi golden rule, where one also integrates
through a collision into future and uses matching of asymptotic states
to identify a product of the completed collision.

In accordance with the causality principle, all time integrals in
(\ref{gf13}) run only over the left part of the time axis, i.e., for
times smaller than $t$. To monitor a close time vicinity, we rearrange
the transport equation (\ref{gf13}) as
\begin{eqnarray}
G^<&=&{1\over 2}(G^R-G^A)\Sigma^<G^A-
{1\over 2}G^R\Sigma^<(G^R-G^A)
\nonumber\\
&+&{1\over 2}G^R\Sigma^<G^R+{1\over 2}G^A\Sigma^<G^A.
\label{ql1}
\end{eqnarray}
In the added term
\begin{equation}
\Xi^<={1\over 2}G^R\Sigma^<G^R+{1\over 2}G^A\Sigma^<G^A,
\label{ql2}
\end{equation}
the time integration runs into future,
\begin{eqnarray}
2\Xi^<(t_1,t_2)&=&\int_{-\infty}^{t_1} dt^\prime
\int_{t_2}^\infty d\hat t
G^R(t_1,t^\prime)\Sigma^<(t^\prime,\hat t)G^R(\hat t,t_2)
\nonumber\\
&+&\int_{-\infty}^{t_2}d\hat t\int_{t_1}^\infty dt^\prime
G^A(t_1,t^\prime)\Sigma^<(t^\prime,\hat t)G^A(\hat t,t_2).
\nonumber\\
&&
\label{ql3}
\end{eqnarray}
The same integration into future also appears in the other contribution
to $G^<$
\begin{equation}
\Lambda^<={1\over 2}(G^R-G^A)\Sigma^<G^A-
{1\over 2}G^R\Sigma^<(G^R-G^A).
\label{ql4}
\end{equation}

On the time diagonal, $t_{1,2}=t$, the time integrations over $t^\prime$
and $\hat t$ in $\Xi^<$ do not overlap, see (\ref{ql3}). The time scale
of the integration is determined by the time scale of $\Sigma^<$ that
can be identified with the quasiparticle formation time.\cite{LKKW91}
Thus $\Xi^<$ is dominated by the short-time (off-pole) contributions.

The time integration in $\Lambda^<$ extends into the future in a way
that reminds Fourier transformation to energy of the asymptotic state,
\begin{equation}
G^R(t,t^\prime;k)-G^A(t,t^\prime;k)
\approx
-i
{\rm e}^{-i\varepsilon_k(t-t^\prime)}
{\rm e}^{-{|t-t^\prime|\over\tau}}.
\label{ql5}
\end{equation}
Thus $\Lambda^<$ is dominated by the long-time (pole) contributions.

\subsection{Pole and off-pole contributions}
Splitting the correlation function $G^<$ into $\Lambda^<$ and $\Xi^<$ is
an ideal starting point to a non-equilibrium modification of the
expansion in small scattering rates. This can be seen in equilibrium,
where one can use the energy representation
\begin{eqnarray}
\lambda^<(\omega,k)&=&f_{FD}(\omega)
{1\over 2}a(\omega,k)^2\gamma(\omega)
\nonumber\\
&\to &f(k)z(k)2\pi\delta(\omega-\varepsilon).
\label{ql6}
\end{eqnarray}
We have used $\sigma^<(\omega)=f_{FD}(\omega)\gamma(\omega)$, and
(\ref{gf11}). The arrow shows a value in the limit of small scattering
rates.

Note that $\lambda^<=f_{FD}{1\over 2}a^2\gamma$, while
$g^<=f_{FD}a$. In the limit of small scattering rate, $\gamma\to 0$,
the spectral function $a$ approaches $\delta$ function as Lorentzian,
while $s={1\over 2}a^2\gamma$ approaches $\delta$ function faster,
\begin{eqnarray}
a(\omega,k)&=&
{\gamma\over(\omega-\epsilon-\sigma)^2+{1\over 4}\gamma^2},
\label{ql7}\\
s(\omega,k)&=&{{1\over 2}\gamma^3\over
\left((\omega-\epsilon-\sigma)^2+{1\over 4}\gamma^2\right)^2}.
\label{ql8}
\end{eqnarray}
In the off-pole region $|\omega-\varepsilon|\gg\gamma$, the spectral
function $a$ has a tail linear in $\gamma$. In the limit of small
scattering rates, $\gamma\to 0$, this tail results in the off-pole
correction, the second term in (\ref{gf12}). The function $s$ in the
off-pole region is proportional to $\gamma^3$, thus its limit is
a pure $\delta$ function without the off-pole term.

The function $\xi^<$ contains the off-pole part. In equilibrium,
\begin{eqnarray}
\xi^<(\omega,k)
&=&
f_{FD}(\omega)
\gamma(\omega)
{\rm Re}\left(\omega-\epsilon-\sigma^R\right)^2
\nonumber\\
&\to &
f_{FD}(\omega)
\gamma(\omega)
{\rm Re}\left(\omega-\epsilon+i0\right)^2.
\label{ql9}
\end{eqnarray}
Due to the off-pole nature of this contribution, the Fermi-Dirac
distribution in (\ref{ql9}) cannot be associated with occupation of the
state $k$.

The comparison with equilibrium shows that the quasiparticle
distribution relates to $\lambda^<$ while $\xi^<$ has to be constructed
indirectly. In the spirit of the BE, we will treat $\lambda^<$ within
a differential transport equation while $\xi^<$ will be turned into a
local functional.

Equilibrium relations (\ref{ql6}-\ref{ql9}) can be easily generalized
to quasiclassical limit. As we have shown, the spectral function $a$
is free of gradients, see (\ref{gf9}) and (\ref{gf11}). The function
$S=A+{1\over 2}G^R\Gamma G^R+{1\over 2}G^A\Gamma G^A$ is also free
of gradients since gradient expansion of symmetric terms like
$G^R\Gamma G^R$ has no gradients. Therefore, similarly to the spectral
function $a$, the function $s$ just follows energy bottom defined by the
field $\phi$
\begin{equation}
s={{1\over 2}\gamma^3\over
\left((\omega-\epsilon-\phi-\sigma)^2+{1\over 4}\gamma^2\right)^2}.
\label{ql10}
\end{equation}
In the limit of small scattering rate, $\gamma\to 0$, the function $s$
reduces to the first term of the spectral function,
\begin{equation}
s=2\pi z\delta(\omega-\varepsilon).
\label{ql11}
\end{equation}

In equilibrium, the pole part $\lambda^<$ is proportional to the
function $s$. Out of equilibrium, we can expect similar behaviour and
introduce local distribution as
\begin{equation}
\lambda^<(\omega,k,r,t)=f_{FD}^{\rm loc}(\omega,k,r,t)s(\omega,k,r,t).
\label{ql12}
\end{equation}
In the limit of small scattering rates, the function $s$ turns to the
$\delta$ function and one can eliminate the energy argument of the
local distribution. In this way we can define the quasiparticle
distribution as the pole of the local distribution. Briefly, in the
limit of small scattering rates the pole part of the correlation
function reads
\begin{equation}
\lambda^<(\omega,k,r,t)=f(k,r,t)2\pi z\delta(\omega-\varepsilon).
\label{ql13}
\end{equation}

The non-equilibrium generalization of the off-pole part $\xi^<$
follows directly from its definition (\ref{ql2}). Since symmetric terms
has no gradient contributions
\begin{equation}
\xi^<=\sigma^<{1\over 2}(g_R^2+g_A^2).
\label{ql14}
\end{equation}
In the limit of small scattering rates
\begin{equation}
\xi^<=\sigma^<{\rm Re}{1\over(\omega-\epsilon-\phi-\sigma+i0)^2}.
\label{ql15}
\end{equation}

\subsection{Boltzmann equation}
The BE is recovered from (\ref{ql4}). First, we turn (\ref{ql4}) into
differential form multiplying it by $G_R^{-1}$ from the l.h.s. and by
$G_A^{-1}$ from the r.h.s. and subtracting the two forms
\begin{eqnarray}
-i&&\left(G_R^{-1}\Lambda^<-\Lambda^<G_A^{-1}\right)
\nonumber\\
&&={1\over 2}
(\Sigma^<A+A\Sigma^<-\Gamma G^A\Sigma^<G^A-G^R\Sigma^<G^R\Gamma ).
\label{ql19}
\end{eqnarray}
By the gradient expansion, this equation simplifies as
\begin{equation}
[\omega-\epsilon-\phi-\sigma,\lambda^<]-
{1\over 2}[\gamma,ag\sigma^<]=\sigma^<s-\gamma\lambda^<,
\label{ql20}
\end{equation}
where $g={\rm Re}g^R$. In the limit of small scattering rates, the
second term on the l.h.s. vanishes. After a substitution of
(\ref{ql13}), equation (\ref{ql20}) turns into equation (\ref{i1}).
For details see Refs.~\onlinecite{SL95,SL94}.

To deal with the transport equation (\ref{i1}), we need: life time
$\tau$, quasiparticle energy $\varepsilon$, wave-function
renormalization $z$, and transport vertex $\sigma^<$. Since all these
functions are related to the self-energy, further progress requires to
specify the self-energy. For elastic scattering on impurities, the
retarded (advanced) self-energy depends only on the retarded (advanced)
propagator $G^R$. The transport vertex depends on both propagators,
$G^R$ and $G^A$, and on the correlation function $G^<$. Propagators
are given by (\ref{gf9}). The correlation function has to be decomposed
into two parts according (\ref{ql1}) which in Wigner's representation
reads
\begin{equation}
g^<=\lambda^<+\xi^<.
\label{ql21}
\end{equation}
The pole part $\lambda^<$ is linked to the quasiparticle distribution
via (\ref{ql13}), the off-pole part $\xi^<$ is self-consistently
evaluated from the transport vertex $\sigma^<$ via (\ref{ql15}).

\section{Averaged T-matrix approximation}
Here we specify the self-energy. As in Paper~I, we assume
non-interacting electrons scattered by neutral impurities. The impurity
acts on electrons by a single-site potential introduced by Koster and
Slater\cite{KS54a,KS54b}. Individual scattering events are treated
within the T-matrix. Formulas for the T-matrix in homogeneous systems
are quite common, \cite{VKE68} our focus will be on gradient
contributions in inhomogeneous systems.

\subsection{Retarded self-energy}
In the Koster-Slater model, an impurity at position $r$ is characterized
by a potential restricted to a single orbital $|r\rangle$ at site $r$,
\begin{equation}
V_r=|r\rangle v\langle r|.
\label{se1}
\end{equation}
Corresponding retarded T-matrix reads
\begin{equation}
T^R_r=V_r+V_rG^RT^R_r.
\label{se2}
\end{equation}
Iterating (\ref{se2}) one can see that the T-matrix is also single-site
function,
\begin{equation}
T^R_r=|r\rangle{v\over 1-\langle r|G^R|r\rangle}\langle r|.
\label{se3}
\end{equation}

The T-matrix does not depend on difference coordinate, therefore its
mixed representation relates only to double-time structure
\begin{equation}
|r\rangle t^R(\omega,t,r)\langle r|=
\int d\tau {\rm e}^{i\omega\tau}
|r\rangle t^R(t+{\tau\over 2},t-{\tau\over 2},r)\langle r|.
\label{se4}
\end{equation}
Similarly, the local element of the propagator also depends only on
energy, not on momentum. In Wigner's representation, the local element
of the retarded propagator $\langle r|G^R(t_1,t_2)|r\rangle\equiv
G^R(r,t_1,r,t_2)\equiv\tilde G^R(r,t_1,t_2)$ transforms into
\begin{equation}
\tilde g^R(\omega,r,t)=\int{dk\over(2\pi)^3}g^R(\omega,k,r,t).
\label{se5}
\end{equation}

The T-matrix in mixed representation reads
\begin{equation}
t^R(\omega,r,t)={v\over 1-v\tilde g^R(\omega,r,t)}.
\label{se6}
\end{equation}
There are no gradient contributions. This can be checked directly by
explicit gradient expansion of (\ref{se3}).

The retarded self-energy is defined as a mean value of the T-matrix
\begin{equation}
\Sigma^R=\int dr c(r)T^R_r,
\label{se7}
\end{equation}
where $c(r)$ is a concentration of impurity per site on site $r$.
This approximation is called the self-consistent averaged T-matrix
approximation (ATA). Unlike in Paper~I, we do not use the subscript
self here since the self-consistent form is a natural starting point
in transport theory. Non-self-consistent values are introduced and
denoted below.

In Wigner's representation (\ref{se7}) reads
\begin{equation}
\sigma^R(\omega,r,t)=c(r)t^R(\omega,r,t).
\label{se8}
\end{equation}

\subsection{Transport vertex}
The transport vertex $\sigma^<$ in the self-consistent ATA depends on
the correlation function $G^<$ as
\begin{equation}
\Sigma^<=\int dr c(r) T^R_rG^<T^A_r,
\label{se9}
\end{equation}
which in Wigner's representation reads
\begin{eqnarray}
\sigma^<&=&ct^R\tilde g^<t^A
\nonumber\\
&+&c{i\over 2}\tilde g^<[t^R,t^A]
\nonumber\\
&+&c{i\over 2}\left(t^A[t^R,\tilde g^<]-t^R[t^A,\tilde g^<]\right).
\label{se10}
\end{eqnarray}
Here,
\begin{equation}
\tilde g^<(\omega,r,t)=\int {dk\over(2\pi)^3}G^<(\omega,k,r,t)
\label{se11}
\end{equation}
is the local element of the correlation function. The Poisson bracket
used in (\ref{se10}) as a short-hand for gradient corrections in general
includes space derivatives combined with derivatives with respect to
momentum, see (\ref{gf7}). In (\ref{se10}), however, none of functions
depends on momentum so that the only gradient contributions come from
time-derivatives.

The transport vertex $\sigma^<$ has three basic components that have
distinguishable physical content. First, there is a non-gradient term
\begin{equation}
\sigma^<_{\rm ng}=ct^R\tilde g^<t^A.
\label{se12}
\end{equation}
Second, there is the term
\begin{equation}
\sigma^<_{\rm P}=c{i\over 2}\tilde g^<[t^R,t^A]
\label{se13}
\end{equation}
which formally brings gradient corrections to the scattering rates.
Below we show that this term vanishes. Third, there are two complex
conjugate terms
\begin{equation}
\sigma^<_\Delta=c{i\over 2}
\left(t^A[t^R,\tilde g^<]-t^R[t^A,\tilde g^<]\right)
\label{se14}
\end{equation}
which contribute if the quasiparticle distribution $f$ has non-zero
gradients. This last term results in non-local corrections of the BE
(\ref{qp12}).

Now the set of equations is complete.

\section{Scattering integrals}
The scattering-out integral $z\gamma_\varepsilon f$ and the non-gradient
part of the scattering-in integral $z\sigma^<_\varepsilon$ are dominant.
They are the only non-zero terms of the BE in the absence of the
perturbing field $\phi$ and their balance determines how the BE is
capable to describe equilibrium.

\subsection{Non-gradient part of the scattering integral}
In the quasiclassical limit, the frequency of
the perturbing field is much smaller than the relaxation time of the
system, i.e., all perturbations are on the long-time scale. Therefore,
the consistency of $z\gamma_\varepsilon f$ and $z\sigma^<_\varepsilon$
is crucial and it has to be checked in very detail.

On the level of non-equilibrium Green's functions, the scattering
in and out are consistent if the self-energy is given by the T-matrix
that obeys the optical theorem. The ATA obeys it. The consistency on
the level of Green's functions does not imply a consistency within
the BE. The scattering-in integral in the BE include additional
approximations (\ref{ql13}) and (\ref{ql15}), while the scattering out
is evaluated without them. Since (\ref{ql13}) and (\ref{ql15}) follow
from the limit of small scattering rates, we have to make corresponding
approximation for the scattering out.

\subsubsection{Scattering out}
The limit of small scattering rates, $\gamma\to 0$, is conveniently
discussed in terms of the non-self-consistent ATA. Indeed, sending
$\sigma^R\to 0$ in propagators $\tilde g^R$ of the self-consistent
T-matrix $t^R$ turns into the non-self-consistent one $t^R_{00}$.

>From (\ref{gf8}) one can see that
\begin{equation}
g^R(\omega)=g^R_{00}(\omega-\phi-\sigma^R),
\label{ng0a}
\end{equation}
where subscript 00 denotes no field and no self-energy in propagator,
[in Paper~I, these non-self-consistent functions are without subscript]
\begin{equation}
g^R_{00}(\omega)={1\over\omega-\epsilon+i0}.
\label{ng0b}
\end{equation}
The self-energy $\sigma^R$ is then expressed in terms of the
non-self-consistent self-energy
\begin{equation}
\sigma^R_{00}=ct^R_{00}=c{v\over 1-v\tilde g^R_{00}}
\label{ng1}
\end{equation}
as
\begin{equation}
\sigma^R(\omega)=\sigma^R_{00}(\omega-\phi-\sigma+{i\over 2}\gamma).
\label{ng2}
\end{equation}
In the limit of small scattering rates we linearize in $\gamma$,
\begin{equation}
\sigma^R(\omega)=\sigma^R_{00}(\omega-\phi-\sigma)+
{i\over 2}\gamma\left.{\partial\sigma^R_{00}\over\partial\omega}
\right|_{\omega-\phi-\sigma}.
\label{ng3}
\end{equation}
The imaginary part of (\ref{ng3}), $\gamma\equiv -2{\rm Im}\sigma^R$,
reads
\begin{equation}
\gamma(\omega)\left(1+{\partial\sigma_{00}\over\partial\omega}
\right)_{\omega-\phi-\sigma}=\gamma_{00}(\omega-\phi-\sigma).
\label{ng4}
\end{equation}

At the pole value $\omega=\varepsilon=\epsilon+\phi+\sigma(\varepsilon)$,
the argument of $\gamma_{00}$ simplifies
\begin{equation}
\gamma(\varepsilon)\left(1+{\partial\sigma_{00}\over\partial\omega}
\right)_{\omega=\epsilon}=\gamma_{00}(\epsilon).
\label{ng4a}
\end{equation}
The factor $1+\left.{\partial\sigma_{00}\over\partial\omega}
\right|_{\omega=\epsilon}$ is just the wave-function renormalization
$z$, see Appendix~A of Paper~I. Indeed, from real part of (\ref{ng3})
one finds
\begin{equation}
\left.{\partial\sigma\over\partial\omega}\right|_{\omega=\varepsilon}=
\left.{\partial\sigma_{00}\over\partial\omega}
\right|_{\omega=\varepsilon-\phi-\sigma}
\left(1-\left.{\partial\sigma\over\partial\omega}
\right|_{\omega=\varepsilon}\right),
\label{ng4b}
\end{equation}
which is equivalent to
\begin{equation}
z={1\over 1-
\left.{\partial\sigma\over\partial\omega}\right|_{\omega=\varepsilon}}=
1+\left.{\partial\sigma_{00}\over\partial\omega}
\right|_{\omega=\epsilon}.
\label{ng4c}
\end{equation}
Accordingly, we have found that the
self-consistent and non-self-consistent approximations are linked via
the wave-function renormalization
\begin{equation}
{1\over\tau}=z\gamma_\varepsilon=\gamma_{00}(\epsilon)
=c(-2){\rm Im}t^R_{00}(\epsilon).
\label{ng5}
\end{equation}

The non-self-consistent T-matrix satisfies the optical theorem
${\rm Im}t^R_{00}=t^R_{00}({\rm Im}\tilde g^R_{00})t^A_{00}$, which
easily follows from (\ref{se6}) and the complex conjugacy of retarded
and advanced functions, $t^A_{00}=(t^R_{00})^*$. The scattering-out
rate thus can be expressed in form of a sum over individual scattering
rates into all accessible finite states
\begin{eqnarray}
z\gamma_{\varepsilon_k}&=&c\left|t^R_{00}(\epsilon_k)\right|^2
\int{dp\over(2\pi)^3}2\pi\delta(\epsilon_k-\epsilon_p).
\label{ng6}
\end{eqnarray}
The $\delta$ function in the r.h.s. results from
$-2{\rm Im}g^R_{00}(\omega)=2\pi\delta(\omega-\epsilon)$. From
(\ref{ng6}) one identifies the scattering rates
\begin{equation}
P_{pk}=c\left|t^R_{00}(\epsilon_k)\right|^2
2\pi\delta(\epsilon_k-\epsilon_p).
\label{ng7}
\end{equation}
Since there are no gradient contributions to the scattering out, the
quantum-mechanical scattering-out rate (\ref{ng7}) is of the same form
as the intuitively expected classical scattering-out integral in the
modified BE (\ref{qp12}).

\subsubsection{Scattering in}
Now we evaluate the non-gradient part $\sigma_{\rm ng}^<$ of the
scattering-in integral (\ref{se12}). We show that it results exactly in
the scattering-in integral from the BE (\ref{qp12}) with the scattering
rates given by (\ref{ng7}). In this way, the consistency of scattering
in and out will be checked.

Substituting $\tilde g^<$ from (\ref{se11}) with $g^<=\lambda^<+\xi^<$
from (\ref{ql13}) and (\ref{ql15}), one gets
\begin{eqnarray}
\sigma^<_{\rm ng}&=&c\left|t^R\right|^2 \tilde g^<
\nonumber\\
&=&c\left|t^R\right|^2\int{dp\over(2\pi)^3}
zf(p)2\pi\delta(\omega-\varepsilon_p)
\nonumber\\
&+&c\left|t^R\right|^2\int{dp\over(2\pi)^3}
\sigma^<_{\rm ng}{\rm Re}
{1\over(\omega-\epsilon_p-\phi-\sigma+i0)^2}.
\label{ng8}
\end{eqnarray}
The first term follows from $\lambda^<$ and is the dominant one. The
second one is the off-pole correction due to $\xi^<$.

The T-matrix in (\ref{ng8}) includes propagators with poles out of
the real energy axis, shifted by $\gamma/2$. In the limit of small
scattering rates, it is advantageous to use the non-self-consistent ATA
as the reference point
\begin{eqnarray}
t^R(\omega)&=&
t^R_{00}(\omega-\phi-\sigma+{i\over 2}\gamma)
\nonumber\\
&=&
t^R_{00}(\omega-\phi-\sigma)+
{i\over 2}
\gamma
\left.
{\partial t^R_{00}\over\partial\omega}
\right|_{\omega-\phi-\sigma}.
\label{ng9}
\end{eqnarray}
>From this approximation one finds that the square of the absolute value
at the pole reads
\begin{eqnarray}
\left|t^R(\varepsilon)\right|^2&=&
\left|t^R_{00}(\epsilon)\right|^2
\nonumber\\
&\times &
\left(
1+\gamma_\varepsilon
{i\over 2}\left(
{1\over t^R_{00}}
{\partial t^R_{00}\over\partial\omega}-
{1\over t^A_{00}}
{\partial t^A_{00}\over\partial\omega}
\right)_{\omega=\epsilon}
\right).
\label{ng10}
\end{eqnarray}
The second term in the bracket in (\ref{ng10}) can be expressed in terms
of the collision delay\cite{F83} (I-25)
\begin{equation}
\Delta_t=\left.{\rm Im}{1\over t^R_{00}}
{\partial t^R_{00}\over\partial\omega}
\right|_{\omega=\epsilon}={i\over 2}\left({1\over t^R_{00}}
{\partial t^R_{00}\over\partial\omega}-{1\over t^A_{00}}
{\partial t^A_{00}\over\partial\omega}\right)_{\omega=\epsilon}
\label{ng10a}
\end{equation}
as
\begin{equation}
\left|t^R(\varepsilon)\right|^2=
\left|t^R_{00}(\epsilon)\right|^2
\left(
1-{\Delta_t\over\tau}{1\over z}
\right).
\label{ng11}
\end{equation}
This relation between the self-consistent and non-self-consistent
T-matrices has a form of virial corrections. These are not, however,
the virial corrections we are looking for. The term
${\Delta_t\over\tau}$ will be cancelled
by the wave-function renormalization and the off-pole contribution.

Let us denote by $\sigma^<_{00}$ the non-self-consistent-like
scattering vertex
\begin{eqnarray}
\sigma^<_{00}(\varepsilon_k)
&=&
c\left|t^R_{00}(\varepsilon_k-\phi-\sigma)\right|^2
\int{dp\over(2\pi)^3}
zf(p)2\pi\delta(\varepsilon_k-\varepsilon_p)
\nonumber\\
&=&
c\left|t^R_{00}(\epsilon_k)\right|^2
\int{dp\over(2\pi)^3}
f(p)2\pi\delta(\epsilon_k-\epsilon_p)
\nonumber\\
&=&
\int{dp\over(2\pi)^3}P_{pk}f(p).
\label{ng12}
\end{eqnarray}
Apparently, $\sigma^<_{00}$ is compatible with the scattering out. To
prove the compatibility of $\sigma_{\rm ng}^<$ with $f\gamma$, we have
to show that all corrections (linear in $\gamma$) following from the
non-self-consistency mutually cancel, i.e., $z\sigma^<_{\rm ng}=
\sigma^<_{00}$.

There are three contributions to the correction
$z\sigma^<_{\rm ng}-\sigma^<_{00}$:
(i) the second term in (\ref{ng8}),
(ii) the ${\Delta_t\over\tau}$ correction to the square of the T-matrix,
and
(iii) the wave-function renormalization in front of
$\sigma^<_\varepsilon$.
Remains to show that the sum of these three corrections vanishes. We
will neglect higher order terms resulting from products of individual
corrections.

First, we rearrange the second term in (\ref{ng8}). To this end we use
that $\sigma^<$ does not depend on the momentum $p$ and move it out of
the integral. The integrand can be rearranged is spirit of Ward's
identities with help of energy derivative as
\begin{equation}
{\rm Re}{1\over(\omega-\epsilon-\phi-\sigma+i0)^2}=-
\left.{\partial g_{00}\over\partial\omega}
{1\over 1-{\partial\sigma\over\partial\omega}}
\right|_{\omega-\phi-\sigma}.
\label{ng14}
\end{equation}
By the integration over $p$, relation (\ref{ng14}) turns into the local
propagator needed in the second term of (\ref{ng8})
\begin{equation}
\int{dp\over(2\pi)^3}{\rm Re}
{1\over(\varepsilon_k-\epsilon_p-\phi-\sigma(\varepsilon_k)+i0)^2}
=-z\left.{\partial\tilde g_{00}\over\partial\omega}
\right|_{\epsilon_k}.
\label{ng16}
\end{equation}
The second term of (\ref{ng8}) thus can be rearranged as
\begin{eqnarray}
&&c\left|t^R\right|^2\int{dp\over(2\pi)^3}\sigma^<_{\rm ng}{\rm Re}
{1\over(\varepsilon_k-\epsilon_p-\phi-\sigma(\varepsilon_k)+i0)^2}
\nonumber\\
&&=-z\sigma^<_{\rm ng}c\left|t^R_{00}(\epsilon_k)\right|^2
\left(1-{\Delta_t\over\tau}{1\over z}\right)
\left.{\partial\tilde g_{00}\over\partial\omega}\right|_{\epsilon_k}.
\label{ng17}
\end{eqnarray}

Now we can collect all terms which contribute to the non-gradient part
of the scattering in
\begin{eqnarray}
z\sigma^<_{\rm ng}(\varepsilon_k)&=&\left(z-{\Delta_t\over\tau}\right)
\sigma^<_{00}(\varepsilon_k)-c|t^R_{00}|^2\left.
{\partial\tilde g_{00}\over\partial\omega}\right|_{\epsilon_k}
z\sigma^<_{\rm ng}(\varepsilon_k)
\nonumber\\
&=&
\left(1+{\partial\sigma_{00}\over\partial\omega}-{\Delta_t\over\tau}-
c|t^R_{00}|^2{\partial\tilde g_{00}\over\partial\omega}
\right)_{\epsilon_k}\sigma^<_{00}(\varepsilon_k)
\nonumber\\
&=&\sigma^<_{00}(\varepsilon_k).
\label{ng18}
\end{eqnarray}
In the last but one step, we have neglected the cross-correction
${\Delta_t\over\tau}\times c|t^R_{00}|^2{\partial\tilde g_{00}\over
\partial\omega}$ and terms quadratic in
$c|t^R_{00}|^2{\partial\tilde g_{00}\over \partial\omega}$ which are
higher order in the limit of small scattering rates. In the last step,
we have used the derived optical theorem (I-B3), [proved also in
Appendix of this paper, see (\ref{vq22})]
\begin{equation}
{\Delta_t\over\tau}=
\left({\partial\sigma_{00}\over\partial\omega}-
c|t^R_{00}|^2{\partial\tilde g_{00}\over\partial\omega}
\right)_{\epsilon_k}.
\label{ng19}
\end{equation}

Briefly, we have shown that in the non-gradient scattering in and out,
the wave-function renormalization $z$ compensates in a consistent
manner. This compensation shows that the off-pole part $\Xi^<$ of the
correlation function is capable to rebound consistently the off-pole
portion of the particle propagation.

The non-gradient parts of scattering integrals
\begin{eqnarray}
z\gamma(\varepsilon_k,r,t)&=&\int{dk\over(2\pi)^3}P_{kp}f(k,r,t),
\label{si1}\\
z\sigma^<_{\rm ng}(\varepsilon_k,r,t)&=&\int{dk\over(2\pi)^3}
P_{pk}f(p,r,t),
\label{si2}
\end{eqnarray}
have scattering rates identical to their non-self-consistent
counterparts
\begin{eqnarray}
P_{pk}&=&c|t^R_{00}(\epsilon_k)|^22\pi\delta(\epsilon_k-\epsilon_p)
\nonumber\\
&=&{1\over\tau}{2\pi^2\over k^2}\delta(|p|-|k|).
\label{si3}
\end{eqnarray}
These non-gradient parts are identical to those used in the intuitive
BE (\ref{qp12a}).

\subsection{Gradient corrections to the scattering rate}
The gradient term of the self-energy $\sigma^<_{\rm P}$ given by
(\ref{se13}) does not include gradients of the quasiparticle
distribution. Accordingly, $\sigma^<_{\rm P}$ can bring corrections
to scattering rates $P_{pk}$ only. In reality $\sigma^<_{\rm P}=0$,
i.e., no correction to scattering rates appears.

For impurity potentials independent from time, $v=const$, the Poisson
bracket in (\ref{se13}) equals zero. Indeed, the both T-matrices depend
on the time $t$ only via the internal potential $\phi$ which always
enters the T-matrix in the form of $\omega-\phi$, i.e.,
$t^{R,A}(\omega,t)=t^{R,A}(\omega-\phi)$. Any two functions of this
property has zero Poisson bracket,
\begin{eqnarray}
[a(\omega-\phi),b(\omega-\phi)]&=&{\partial a\over\partial\omega}
{\partial b\over\partial t}-{\partial a\over\partial t}
{\partial b\over\partial\omega}
\nonumber\\
&=&-{\partial a\over\partial\omega}
{\partial b\over\partial\omega}{\partial\phi\over\partial t}+
{\partial a\over\partial\omega}{\partial\phi\over\partial t}
{\partial b\over\partial\omega}
\nonumber\\
&=&0.
\label{gc3}
\end{eqnarray}
Accordingly,
\begin{equation}
z\sigma^<_{\rm P}=0.
\label{gc11}
\end{equation}

Briefly, amplitudes of scattering rates are not modified by gradient
corrections for time-independent impurity potentials. This is in
agreement with the intuitive expectation used in Paper~I.

\subsection{Gradient corrections to scattering in}
The gradient correction $\sigma^<_\Delta$ brings true non-local
contributions to the scattering-in integral. Via gradients of the
local correlation function $\tilde g^<$, the correction
$\sigma^<_\Delta$ depends on gradients of the quasiparticle
distribution. Here we show that from $\sigma^<_\Delta$ one
recovers virial corrections in form discussed in Paper~I.

First, we write (\ref{se14}) in explicit form
\begin{eqnarray}
\sigma^<_\Delta
&=&ct^Rt^A{i\over 2}\left({1\over t^R}[t^R,\tilde g^<]-
{1\over t^A}[t^A,\tilde g^<]\right)
\nonumber\\
&=&-c|t^R|^2{\rm Im}{1\over t^R}[t^R,\tilde g^<]
\nonumber\\
&=&-c|t^R|^2{\rm Im}{1\over t^R}
\left({\partial t^R\over\partial\omega}
{\partial\over\partial t}-
{\partial t^R\over\partial t}
{\partial\over\partial\omega^<}\right)\tilde g^<.
\label{gc12}
\end{eqnarray}
Now, we substitute the dominant part of the local correlation function
(the off-pole part leads to higher order contribution in small $\gamma$)
into (\ref{se14})
\begin{equation}
g^<(\omega,r,t)
\approx\int{dp\over(2\pi)^3}f(p,r,t)2\pi z\delta(\omega-\varepsilon)
\label{gc14}
\end{equation}
and interchange derivatives with the momentum integral
\begin{eqnarray}
\sigma^<_\Delta
&=&-c|t^R|^2\int{dp\over(2\pi)^3}{\rm Im}{1\over t^R}
\nonumber\\
&\times & \left({\partial t^R\over\partial\omega}
{\partial\over\partial t}-{\partial t^R\over\partial t}
{\partial\over\partial\omega}\right)
f(p)2\pi z\delta(\omega-\varepsilon_p).
\label{gc15}
\end{eqnarray}
The function $z\delta(\omega-\varepsilon)$ depends on time only via
$\omega-\phi$ [$z\delta(\omega-\varepsilon)=\delta(\omega-\epsilon-\phi-
\sigma)$], therefore according to (\ref{gc3}), $\left({\partial t^R\over
\partial\omega}{\partial\over\partial t}-{\partial t^R\over\partial t}
{\partial\over\partial\omega}\right)2\pi z\delta(\omega-\varepsilon_p)=
0$. The gradient correction $\sigma^<_\Delta$ thus depends exclusively
on gradients of the distribution function,
\begin{equation}
\sigma^<_\Delta
=-c|t^R|^2\int{dp\over(2\pi)^3}2\pi z\delta(\omega-\varepsilon_p)
{\rm Im}{1\over t^R}{\partial t^R\over\partial\omega}
{\partial\over\partial t}f(p).
\label{gc15a}
\end{equation}
Finally, we simplify this gradient correction with help of (\ref{ng11}),
scattering rate (\ref{ng7}) and the collision delay (\ref{ng10a}) as
\begin{eqnarray}
z\sigma^<_\Delta &=&-\left(z-{\Delta_t\over\tau}\right)
c|t_{00}^R|^2\int{dp\over(2\pi)^3}2\pi z\delta(\omega-\varepsilon)
\Delta_t{\partial\over\partial t}f(p)
\nonumber\\
&=&-\int{dp\over(2\pi)^3}P_{pk}{\partial f\over\partial t}\Delta_t.
\label{gc17}
\end{eqnarray}
In the second line, we have neglected higher order terms using the
approximation $z-\Delta_t/\tau\approx 1$.

This gradient correction to the scattering integrals has the form of
virial corrections, the last term of (\ref{qp12a}). In particular, no
other gradient than the time derivative of the quasiparticle
distribution appears, and this time derivative is weighted with the
collision delay and the scattering rate.

\section{Recovering the intuitive transport equation}
Now we can put together elements of the transport equation (\ref{i1})
and reconstruct (\ref{qp12}).

\subsection{Drift part of the transport equation}
The velocity results from momentum derivative of (\ref{gf10}) as
\begin{eqnarray}
{\partial\varepsilon\over\partial k}&=&
{\partial\epsilon\over\partial k}+
\left.{\partial\sigma(\omega)\over\partial\omega}
\right|_{\omega=\varepsilon}
{\partial\varepsilon\over\partial k}
\nonumber\\
&=&{1\over1-\left.{\partial\sigma(\omega)\over\partial\omega}
\right|_{\omega=\varepsilon}}{\partial\epsilon\over\partial k}.
\label{gf11xx}
\end{eqnarray}
For parabolic kinetic
energy, $\epsilon={k^2\over 2m}$, the quasiparticle velocity gains
the form used in the BE (\ref{qp12}),
\begin{equation}
{\partial\varepsilon\over\partial k}=z{k\over m}.
\label{gf13x}
\end{equation}

The force acting on quasiparticle is also found from the quasiparticle
energy,
\begin{equation}
{\partial\varepsilon\over\partial r}=
{\partial\phi\over\partial r}+{\partial\sigma\over\partial r}+
{\partial\sigma\over\partial\omega}
{\partial\varepsilon\over\partial r}.
\label{gf14}
\end{equation}
For homogeneous distribution of impurities, the self-energy depends on
coordinate $r$ only via potential $\phi$. Since the potential $\phi$ can
be viewed as a local shift of initial of energies, one finds that the
self-energy $\sigma$ relates to the self-energy $\sigma_{\phi=0}$ in
the absence of the field $\phi$ as
\begin{equation}
\sigma(\omega)=\sigma_{\phi=0}(\omega-\phi).
\label{gf15}
\end{equation}
>From (\ref{gf15}) one finds that the second term of the force
(\ref{gf14}) reads
\begin{equation}
{\partial\sigma\over\partial r}=-
{\partial\sigma\over\partial\omega}
{\partial\phi\over\partial r}.
\label{gf16}
\end{equation}
Using (\ref{gf16}) in (\ref{gf14}), one finds that the force has
no renormalization
\begin{equation}
{\partial\varepsilon\over\partial r}=
{\partial\phi\over\partial r},
\label{gf17}
\end{equation}
which is the form of the force in the BE (\ref{qp12}). Thus drift terms
of (\ref{i1}) reduce to drift terms of (\ref{qp12}).

\subsection{Scattering integral with virial corrections}
To obtain the expected form (\ref{qp12}), we use that within linear
approximation
\begin{equation}
a(x)+{\partial a\over\partial x}\Delta=a(x+\Delta).
\label{gc16}
\end{equation}
Thus the non-gradient and the gradient scattering-in contributions can be
collected into a compact expression
\begin{eqnarray}
z\sigma^<_\varepsilon &=& z\sigma^<_{\rm ng}(\varepsilon_k)+
z\sigma^<_\Delta(\varepsilon_k)
\nonumber\\
&=&\int{dp\over(2\pi)^3}P_{pk}\left(f-
{\partial f\over\partial t}\Delta_t \right)
\nonumber\\
&=&\int{dp\over(2\pi)^3}P_{pk}f(p,r,t-\Delta_t)
\nonumber\\
&=&{1\over\tau}{2\pi^2\over k^2}
\int{dp\over(2\pi)^3}\delta(|p|-|k|)f(p,r,t-\Delta_t).
\label{gc17a}
\end{eqnarray}
One can see that the scattering in has the form expected from classical
assumptions.

Substituting (\ref{ng5}) for the scattering out, (\ref{gc17a}) for
the scattering in, (\ref{gf13x}) for the quasiparticle velocity, and
(\ref{gf17}) for the accelerating force
into the asymptotic equation (\ref{i1}), the intuitive modification of
the BE (\ref{qp12}) is recovered from quantum statistics.

\section{Observables}
To recover the transport equation (\ref{qp12}) was our major task.
With respect to applications, one has also to find relation between
quasiparticle distribution $f$ and observables. As already shown in
Paper~I, these relations include quasiparticle and virial corrections.
In Paper~I we have discussed only those observables that can be
identified from the transport equation via conservation laws. Here
we extend our treatment to a general single-particle observable.

All single-electron observables can be expressed in terms of the
reduced density matrix (Wigner's distribution function)
\begin{equation}
\rho(k,r,t)=\int {d\omega\over 2\pi}G^<(\omega,k,r,t).
\label{o1}
\end{equation}
>From the decomposition $g^<=\lambda^<+\xi^<$, where $\lambda^<$ and
$\xi^<$ are given by (\ref{ql13}) and (\ref{ql15}), respectively, one
finds the reduced density matrix as
\begin{equation}
\rho=zf-\int{d\omega\over 2\pi}{\wp\over\omega-\varepsilon}
{\partial\sigma^<\over\partial\omega}.
\label{o2}
\end{equation}
Note that the transport vertex $\sigma^<$ in (\ref{o2}) does not
enters the reduced density only by its pole value $\sigma^<_\varepsilon$
but the full energy dependence has to be maintained. Since the second
term is already an off-pole correction, the correlation function of the
self-energy is used only in its lowest approximation $\sigma^<=
\sigma^<_{00}$.

The formula (\ref{o2}) has no explicit gradient terms, however, there
are gradient contributions hidden in the transport vertex $\sigma^<$.
There is a question whether one should keep these gradient corrections
in the off-pole part of formula (\ref{o2}) or not. A general answer is
not clear to us. On the other hand, with respect to conservation laws
that test a consistency of observables with the transport equation
(\ref{qp12}), the gradient contributions can be neglected. The transport
equation does not provide observables but only their time or space
derivatives, see e.g. the equation of continuity (I-C2), therefore any
gradient contribution to observables enter the conservation law via
second derivatives that are neglected within the quasiclassical limit.
Accordingly, we neglect gradient corrections to $\sigma^<$ in
(\ref{o2}).

\subsection{Local density of electrons}
In Paper~I we have derived the local density of electrons $n$ from
the transport equation, see (I-5). For scattering on the Koster-Slater
impurities it was found that the correlated density $n_{\rm corr}$ is
determined by the ratio of the collision delay to the lifetime, see
(I-29). Here we recover (I-29) directly from (\ref{o2}).

The local density of quasiparticles reads
\begin{equation}
n_{\rm free}
=
\int{dk\over(2\pi)^3} f(k).
\label{o3}
\end{equation}
The local density of electrons is given by the integral from $\rho$ over
momentum
\begin{equation}
n=
\int{dk\over(2\pi)^3}
\rho .
\label{o4}
\end{equation}
>From (\ref{o4}) and (\ref{o2}) one finds
\begin{eqnarray}
n
&=&
\int{dk\over(2\pi)^3} f(k)
\nonumber\\
&+&
\int{dk\over(2\pi)^3}
\left.{\partial\sigma_{00}\over\partial\omega}
\right|_{\omega=\epsilon_k} f(k)
\nonumber\\
&+&
\int{dk\over(2\pi)^3}
\int{d\omega\over 2\pi}
{\wp\over\omega-\varepsilon_k}
{\partial\sigma^<_{00}(\omega)\over\partial\omega},
\label{o5}
\end{eqnarray}
where we have used (\ref{ng4c}) for the wave-function renormalization
$z$. The first term is the quasiparticle density $n_{\rm free}$, the
second and the third terms are the correlated density
\begin{eqnarray}
n_{\rm corr}
&=&
\int{dk\over(2\pi)^3}
\left.{\partial\sigma_{00}\over\partial\omega}
\right|_{\omega=\epsilon_k} f(k)
\nonumber\\
&+&
\int{dk\over(2\pi)^3}
\int{d\omega\over 2\pi}
{\wp\over\omega-\varepsilon_k}
{\partial\sigma^<_{00}(\omega)\over\partial\omega}.
\label{o6}
\end{eqnarray}
In the second term we perform the integration by parts
\begin{equation}
{\partial\over\partial\omega}
{\wp\over\omega-\varepsilon}=-{\rm Re}
{1\over(\omega-\varepsilon+i0)^2},
\label{o7}
\end{equation}
and substitute for the self-energy $\sigma^<_{00}$
from (\ref{ng12})
\begin{eqnarray}
n_{\rm corr}
&=&
\int{dk\over(2\pi)^3}f(k)
\left.
{\partial\sigma_{00}\over\partial\omega}
\right|_{\omega=\varepsilon_k}
\nonumber\\
&+&
\int{dk\over(2\pi)^3}{d\omega\over 2\pi}
{\rm Re}{1\over(\omega-\varepsilon_k+i0)^2}
c
\left|t^R_{00}(\omega-\sigma-\phi)
\right|^2
\nonumber\\
&\times &
\int{dp\over(2\pi)^3}
zf(p)2\pi\delta(\omega-\varepsilon_p).
\label{o8}
\end{eqnarray}
Due to the energy-conserving $\delta$ function, the energy $\omega$ can be
easily integrated out. The wave-function renormalization $z$ under the $p$
integral can be omitted as higher order in the limit of small scattering
rates,
\begin{eqnarray}
n_{\rm corr}
&=&
\int{dk\over(2\pi)^3}f(k)
\left.
{\partial\sigma_{00}\over\partial\omega}
\right|_{\omega=\epsilon_k}
\nonumber\\
&+&
c\int{dk\over(2\pi)^3}{dp\over(2\pi)^3}f(p)
{\rm Re}{1\over(\epsilon_p-\epsilon_k+i0)^2}
\left|t^R_{00}(\epsilon_p)\right|^2.
\nonumber\\
\label{o9}
\end{eqnarray}
We have used (\ref{gf10}) to simplify energy arguments. Now we can
integrate over momentum $k$ using non-self-consistent version of
(\ref{ng16}),
\begin{eqnarray}
n_{\rm corr}
&=&
\int{dk\over(2\pi)^3}f(k)
\left.
{\partial\sigma_{00}\over\partial\omega}
\right|_{\omega=\epsilon_k}
\nonumber\\
&-&
c\int{dp\over(2\pi)^3}f(p)
\left.{\partial\tilde g_{00}\over\partial\omega}
\right|_{\omega=\epsilon_p}
\left|t^R_{00}(\epsilon_p)\right|^2.
\label{o10}
\end{eqnarray}

The two terms can be joint. In the first term we rename the integration
variable $k$ to $p$, so that both terms will have the same name of the
momentum argument of the distribution function $f$,
\begin{equation}
n_{\rm corr}=
\int{dp\over(2\pi)^3}f(p)
\left({\partial\sigma_{00}\over\partial\omega}-
c\left|t^R_{00}\right|^2{\partial\tilde g_{00}\over\partial\omega}
\right)_{\omega=\epsilon_p}.
\label{o11}
\end{equation}
Finally, we apply the derived optical theorem (\ref{ng19}) to recover
the relation (I-29)
\begin{equation}
n_{\rm corr}=\int{dp\over(2\pi)^3}f(p){\Delta_t\over\tau}.
\label{o12}
\end{equation}
The total density $n=n_{\rm free}+n_{\rm corr}$ resulting from the
reduced density matrix (\ref{o2}) is thus identical to the one obtained
from the transport equation via the equation of continuity,
\begin{equation}
n=\int{dp\over(2\pi)^3}f(p)
\left(1+{\Delta_t\over\tau}\right),
\label{o12a}
\end{equation}
which is Eq~(I-48).
Briefly, with respect to the electron density, the approximative
functional for the reduced density matrix (\ref{o2}) is consistent with
approximations in the transport equation (\ref{qp12}).

\subsection{Local density of current}
The particle current is one of the quantities most often evaluated from
the BE. Here we show that for the Koster-Slater impurities there are no
explicit virial corrections in the functional for current.

A general formula for the current is
\begin{equation}
j=\int{dk\over(2\pi)^3}
{\partial\epsilon\over\partial k}\rho
=\int{dk\over(2\pi)^3}
{k\over m}\rho.
\label{k22}
\end{equation}
Now we substitute for $\rho$ from (\ref{o2})
\begin{eqnarray}
j
&=&
\int{dk\over(2\pi)^3}
{k\over m}zf
\nonumber\\
&+&
\int{dk\over(2\pi)^3}{d\omega\over 2\pi}
{k\over m}
{\rm Re}
{\sigma^<(\omega)\over(\omega-\varepsilon_k+i0)^2}
\nonumber\\
&=&
\int{dk\over(2\pi)^3}
{k\over m}zf .
\label{k23}
\end{eqnarray}
The second term in the first line is zero because its integrand is an
odd function of momentum $k$. The current is thus identical to (I-41)
found from the equation of continuity.

\subsection{Local density of energy}
Energy of the system is not a single-electron observable.
Although electrons do not mutually interact, the energy of the system
cannot be evaluated from the reduced density matrix $\rho$. This is
because electrons are correlated with impurities of unknown positions.
Similarly to interacting systems, the energy has to be evaluated
directly from the correlation function $g^<$,
\begin{equation}
E=\int{d\omega\over 2\pi}{dk\over(2\pi)^3}
\omega g^<(\omega,k).
\label{k10}
\end{equation}
To express the energy in terms of quasiparticle distribution, we use
$g^<=\lambda^<+\xi^<$ with $\lambda^<$ from (\ref{ql13}) and $\xi^<$
from (\ref{ql15}),
\begin{eqnarray}
E&=&\int{d\omega\over 2\pi}{dk\over(2\pi)^3}
\omega
\left(
\left(
1-{\partial\sigma\over\partial\omega}
\right)^{-1}
f(k)2\pi\delta(\omega-\varepsilon_k)
\right.
\nonumber\\
&+&
c|t^R(\omega)|^2
\int{dp\over(2\pi)^3}
f(p)2\pi\delta(\omega-\varepsilon_p)
\nonumber\\
&\times &
\left.
{\rm Re}
{1\over(\omega-\varepsilon_k+i0)^2}
\right).
\label{k11}
\end{eqnarray}
Now we integrate out the energy $\omega$ and interchange the order of
integrations over momentum in the off-shell term,
\begin{eqnarray}
E&=&\int{dk\over(2\pi)^3}
\varepsilon_k f(k)
+
\int{dk\over(2\pi)^3}
\varepsilon_k f(k)
\left.{\partial\sigma_{00}\over\partial\omega}
\right|_{\omega=\epsilon_k}
\nonumber\\
&+&
c\int{dp\over(2\pi)^3}
|t^R_{00}(\epsilon_p)|^2
f(p)
\nonumber\\
&\times &
\int{dk\over(2\pi)^3}
{\rm Re}
{1\over(\epsilon_p-\epsilon_k+i0)^2}.
\label{k12}
\end{eqnarray}
Similarly as in the case of the electron density, we join the second
and the third terms, evaluate the integration over momentum $k$, and
apply the derived optical theorem (\ref{ng19}) to express the local
energy in terms of the collision delay
\begin{equation}
E=\int{dk\over(2\pi)^3}
\varepsilon_k f(k)
\left(1+{\Delta_t\over\tau}\right).
\label{k12s}
\end{equation}
[In notation of Paper~I, Eq.~(I-42), the quasiparticle energy does not
include the potential $\phi$ that is in (I-42) explicitly added.]
Thermodynamical consistency of the energy conservation and correlated
density is shown in Appendix~C of Paper~I.

The BE (\ref{qp12}) with subsidiary relations (\ref{o12a}), (\ref{k23})
and (\ref{k12s}) form a basic set of equations that cover most of
traditional applications of the BE. In all these equations, the virial
corrections can be included with help of collision delay. We remind
that this simplicity follows in part from the simplicity of the
scattering by Koster-Slater impurities.

\section{Summary}
The intuitive modification of the BE, Eq. (\ref{qp12}), has been
recovered from non-equilibrium statistics. To this end we have
employed non-equilibrium Green's functions within which we made the
quasiclassical limit and the limit of small scattering rates. These
two limits are fully sufficient, i.e., no unjustified approximations
need to be made. The non-local form of the scattering integral in
the intuitive BE has been obtained by unification of non-gradient
and gradient contributions.

Single-electron observables as functionals of the quasiparticle
distribution are provided by the reduced density matrix that in the
limit of small scattering rates has form (\ref{o2}). It was shown
that (\ref{o2}) is consistent with the transport equation (\ref{qp12})
leading to the correct equation of continuity discussed already in
Paper~I. The density of energy, which does not belong to single-electron
observables, has been treated separately.

Presented theory has four general features that can be transferred
to more general models. First, one needs sufficiently complex
self-energy, the recommended one is based on the T-matrix which
guarantees number of identities related to the optical theorem.
Second, for small scattering rate, one can use the procedure of
Refs.~\onlinecite{SL95,SL94} to derive a quasiclassical transport
equation for quasiparticles. Resulting transport equation includes the
quasiparticle and the virial corrections. The virial corrections are,
however, in a form of gradient contributions to the scattering integral.
Third, the virial corrections are rearranged to the semi-classical form
when one recollects non-gradient and gradient terms on the scattering
integral using logarithmic derivatives. Fourth, all logarithmic
derivatives should be defined from the T-matrix, i.e., from the
scattering phase shift. These logarithmic derivatives have natural
interpretations like the collision delay discussed here. Authors of
this paper have implemented presented approach to nuclear matter,
see Ref.~\onlinecite{SLM96}.

On the other hand, the discussed scattering on neutral impurities allows
for a number of simplifications that are not available for more general
scattering mechanisms. First, the self-energy and the transport vertex
are independent of momentum which allows us to employ shifts in the
complex plane with help of which one can conveniently express
self-consistent quantities by their non-self-consistent counterparts.
Second, the lack of momentum dependence leads to the lack of space
non-localities of the scattering integral, therefore all virial
corrections are described by the collision delay. Third, the momentum
independence reflects that there is only a single scattering channel of
s-symmetry. In general, different scattering channels have different
collision delays, in the case of neutral impurities there is only a
single collision delay what simplifies appreciably all related formulas.
Fourth, due to the time-independence of the impurity potential, there are
no gradient corrections to the scattering rate. Fifth, this
time-independence also simplifies the energy conservation in collisions.
Sixth, due to the absence of the dynamics of impurities, the virial
corrections appear only in the scattering in. Here we have selected this
simple scattering on neutral impurities to have free hands to focus on
details of the method.

We have not discussed here consequences and interpretation of the
virial corrections as it has been already done within the intuitive
approach in Paper~I. Our aim here is to confirm validity of this
intuitive approach.

\acknowledgments
This work was supported from the Grant Agency of the Czech Republic
under contract Nr. 202960098, the BMBF (Germany) under contract Nr.
06R0745(0), and the EC Human Capital and Mobility Programme.
\appendix
\section{Virial versus quasiparticle corrections in equilibrium}
The quasiparticle and the virial corrections enter the BE in different
ways. From this point of view, they represent independent corrections
that can be treated separately. On the other hand, for scattering by
resonant levels, both corrections are of the same magnitude as it is
demonstrated in Fig.~5 of Paper~I. Striking similarity of their
magnitudes raises a question up to what extent these two corrections
are independent. To answer this question we briefly discuss equilibrium
where one can benefit from well developed theory of quasiparticle and
virial corrections based on Green's functions.
\cite{KKL84,ZS85,SRS90,MR95}

In equilibrium, the local density of electrons is given by the
spectral function as
\begin{equation}
n=\int{d\omega\over 2\pi}f_{FD}(\omega)\int{dk\over(2\pi)^3}a(\omega,k).
\label{vq2}
\end{equation}
In the limit of small scattering rates, the spectral function will
be substituted from (\ref{gf12}) with $\phi=0$ and lowest order
approximation of the scattering rate, $\gamma=\gamma_{00}$.

In the limit of small scattering rates we can easily separate the
quasiparticle contribution to the local density
\begin{eqnarray}
n_{\rm pole}&=&\int{d\omega\over 2\pi}f_{FD}(\omega)
\int{dk\over(2\pi)^3}z2\pi\delta(\omega-\varepsilon_k)
\nonumber\\
&=&\int{dk\over(2\pi)^3}zf(k),
\label{vq8}
\end{eqnarray}
where $f(k)\equiv f_{FD}(\varepsilon_k)$, and the background
contribution
\begin{equation}
n_{\rm off}=\int{d\omega\over 2\pi}f_{FD}(\omega)
\int{dk\over(2\pi)^3}{\rm Re}{\gamma_{00}(\omega)\over
(\omega-\epsilon_k+i0)^2}.
\label{vq9}
\end{equation}
Using (\ref{ng4c}), $n_{\rm pole}$ can be split into the free part
(\ref{o3}) and wave-function renormalization reduction
\begin{equation}
n_{\rm pole}=n_{\rm free}+n_{\rm wfr},
\label{vq10a}
\end{equation}
where
\begin{equation}
n_{\rm wfr}=\int{dk\over(2\pi)^3}f(k)\left.
{\partial\sigma_{00}\over\partial\omega}
\right|_{\omega=\epsilon_k}.
\label{vq10}
\end{equation}
>From decompositions of the density, $n=n_{\rm pole}+n_{\rm off}$ and
$n=n_{\rm free}+n_{\rm corr}$, and (\ref{vq10a}) the correlated density
results as a sum of the off-pole and wave-function renormalization parts
\begin{equation}
n_{\rm corr}=n_{\rm off}+n_{\rm wfr}.
\label{vq11}
\end{equation}
Using the Kramers-Kr\"onig relation for the real part of the self-energy
\begin{equation}
\sigma_{00}(\omega)={\rm Re}\int{dE\over 2\pi}{1\over E-\omega+i0},
\label{vq11x}
\end{equation}
in (\ref{vq10}), one finds that the correlated density reads
\begin{equation}
n_{\rm corr}=\int{d\omega\over 2\pi}{dk\over(2\pi)^3}
\left(f_{FD}(\omega)-f(k)\right)
{\rm Re}{\gamma_{00}(\omega)\over(\omega-\epsilon_k+i0)^2},
\label{vq11a}
\end{equation}
where the terms weighted by $f_{FD}(\omega)$ and $f(k)$ results from
$n_{\rm off}$ and $n_{\rm wfr}$, respectively. Apparently, there is a
partial compensation of these contributions to $n_{\rm corr}$.

The quantum-mechanical expression (\ref{vq11a}) can be also given the
form of the semi-classical formula (\ref{o12}). To this end we
reorganize $n_{\rm off}$ starting from (\ref{vq9}),
\begin{eqnarray}
n_{\rm off}&=&
\int{d\omega\over 2\pi}f_{FD}(\omega)\gamma_{00}(\omega)
{\rm Re}\int{dk\over(2\pi)^3}{1\over(\omega-\epsilon_k+i0)^2}
\nonumber\\
&=&-\int{d\omega\over 2\pi}f_{FD}(\omega)\gamma_{00}(\omega)
{\partial\over\partial\omega}{\rm Re}\int{dk\over(2\pi)^3}
{1\over\omega-\epsilon_k+i0}
\nonumber\\
&=&-\int{d\omega\over 2\pi}f_{FD}(\omega)\gamma_{00}(\omega)
{\partial\tilde g_{00}\over\partial\omega}.
\label{vq12}
\end{eqnarray}
In the last line we have used that the integral over momentum above
defines a local element of Green's function.

To evaluate $\gamma_{00}$, we use
\begin{equation}
\gamma_{00}(\omega)=c(-2){\rm Im}t^R_{00}(\omega),
\label{vq13}
\end{equation}
and the optical theorem [that follows from non-self-consistent form of
(\ref{se6})]
\begin{equation}
{\rm Im}t^R_{00}=|t^R|^2_{00}{\rm Im}\tilde g_{00}^R .
\label{vq14}
\end{equation}
If we express the local density of states in terms of the momentum
integration
\begin{equation}-2{\rm Im}\tilde g_{00}^R(\omega)=
\int{dp\over(2\pi)^3}
2\pi\delta(\omega-\epsilon_p),
\label{vq15}
\end{equation}
the off-pole contribution can be expressed in terms of the quasiparticle
distribution,
\begin{eqnarray}
n_{\rm off}
&=&-c\int{dp\over(2\pi)^3}{d\omega\over 2\pi}f_{FD}(\omega)
|t^R_{00}(\omega)|^2 2\pi\delta(\omega-\epsilon_p)
{\partial\tilde g_{00}\over\partial\omega}
\nonumber\\
&=&-c\int{dp\over(2\pi)^3}f(p)|t^R_{00}(\epsilon_p)|^2
\left.{\partial\tilde g_{00}\over\partial\omega}
\right|_{\omega=\epsilon_p}.
\label{vq16}
\end{eqnarray}
Finally, we substitute (\ref{vq16}) into (\ref{vq11})
\begin{equation}
n_{\rm corr}=
\int{dp\over(2\pi)^3}
f(p)\left({\partial\sigma_{00}\over\partial\omega}-c|t^R_{00}|^2
{\partial\tilde g_{00}\over\partial\omega}\right)_{\omega=\epsilon_p}.
\label{vq17}
\end{equation}
Formula (\ref{vq17}) is identical to the semi-classical expression
(\ref{o12}). To prove this claim we employ the derived optical theorem.

It is advantageous to start from (\ref{o12}). First, we reorganize
the ratio of the collision delay given by (\ref{ng10a}) to the lifetime
given by (\ref{ng5}) as
${1\over\tau}=2c{\rm Im}t^R_{00}=ic(t^R_{00}-t^A_{00})$ as
\begin{eqnarray}
{\Delta_t\over\tau}
&=&ic(t^R-t^A){1\over 2i}\left({1\over t^R_{00}}
{\partial t^R_{00}\over\partial\omega}-
{1\over t^A_{00}}{\partial t^A_{00}\over\partial\omega}\right)
\nonumber\\
&=&{c\over 2}{\partial\over\partial\omega}(t^R_{00}+t^A_{00})-
{c\over 2}\left({t^A_{00}\over t^R_{00}}
{\partial t^R_{00}\over\partial\omega}+
{t^R_{00}\over t^A_{00}}{\partial t^A_{00}\over\partial\omega}\right)
\nonumber\\
&=&{\partial\sigma_{00}\over\partial\omega}-{c\over 2}\left(
{t^A_{00}\over t^R_{00}}{\partial t^R_{00}\over\partial\omega}+
{t^R_{00}\over t^A_{00}}{\partial t^A_{00}\over\partial\omega}\right).
\label{vq20}
\end{eqnarray}
>From a non-self-consistent form of (\ref{se6}) one finds
\begin{equation}
{\partial t^R_{00}\over\partial\omega}=
{t^R_{00}}^2{\partial\tilde g^R\over\partial\omega},
\label{vq21}
\end{equation}
which substituted into (\ref{vq20}) provides the derived optical theorem
\begin{equation}
{\Delta_t\over\tau}={\partial\sigma_{00}\over\partial\omega}-
c|t^R_{00}|^2
{\partial\tilde g_{00}\over\partial\omega}.
\label{vq22}
\end{equation}
The l.h.s is exactly the bracket in (\ref{vq17}). Thus the semi-classical
formula (\ref{o12}) is equivalent to the quantum-mechanical formula
(\ref{vq17}).

To summarize this appendix we would like to remind the most important
point. Within Green's function, virial and quasiparticle corrections
enter the density together in an unresolved form, but the optical
theorem and the derived optical theorem can be used to separate them and
express virial corrections in terms of the collision delay.

\end{document}